\documentclass[aps,prl,lengthcheck,superscriptaddress]{revtex4-1}

\usepackage{amssymb}
\usepackage{amsmath}
\usepackage{graphicx}

\begin{document}


\title{Current-dependent exchange-correlation potential for non-local absorption in quantum hydrodynamic theory}


\author{Cristian Cirac\`i}
\email[]{cristian.ciraci@iit.it}
\affiliation{Center for Biomolecular Nanotechnologies (CBN), Istituto Italiano di Tecnologia (IIT), Via Barsanti 14, 73010 Arnesano (LE), Italy.}


\date{\today}

\begin{abstract}

The quantum hydrodynamic theory is a promising method for describing microscopic details of macroscopic systems.
The hydrodynamic equation can be directly obtained from a single particle Kohn-Sham equation that includes the contribution of an external vector potential.
This derivation allows to straightforwardly incorporate in the hydrodynamic equation an exchange-correlation viscoelastic term, so that broadening of collective excitation can be taken into account, as well as a correction to the plasmon dispersion.
The result is an accurate self-consistent and computationally efficient hydrodynamic description of the free electron gas.
A very accurate agreement with full quantum calculations is shown.
\end{abstract}


\maketitle

Plasmonic nano-systems have enabled the possibility to macroscopically probe effects that are generally confined to the microscopic realm.
Non-local electron response\cite{Raza:2011io,FernandezDominguez:2012eg,Toscano:2012fh,Ciraci:2013jt,Yan:2016gz} and tunneling effects\cite{Zuloaga:2009gm,Zhu:2016fn,Teperik:2013dd} have be experimentally observed\cite{Ciraci:2012fp,Savage:2012by,Scholl:2013ge,Ciraci:2014jv,Hajisalem:2014cr,Lin:2015et} in plasmonic systems characterized by sub-nanometer gaps.
The advances of fabrication techniques is allowing to control features of such systems at the angstrom scale\cite{Lassiter:2014gs,Kheifets:2014hq,Chikkaraddy:6ja}.
In this context, it becomes very important to develop simulation techniques that are able to take into account quantum microscopic features at the scale of billions of atoms.
Density functional theory (DFT) methods are generally unsuitable because their computational cost grows as fast as $O(N_e^3)$, although a slight improvement can be obtained by using orbital-free techniques\cite{Xiang:2014cz,Xiang:2016cw}.
An alternative but drastic approach is to use a simple linearized Thomas-Fermi (TF) hydrodynamic theory\cite{Ciraci:2013dz,Raza:2015ef}, in which the microscopic problem is only solved for the macroscopic observable quantities, such as, the electron density $n$ and the density of currents ${\bf J}$. 
This simple approach however completely neglects essential effects such as electron spill-out and quantum tunneling. 
A much better description of microscopic details of large systems is given by the quantum hydrodynamic theory (QHT), in which the simple TF kinetic energy of the free-electron gas is modified by including non-local contributions. 
A self-consistent approach based on the QHT was proposed by Toscano \textit{et al.}, who applied it to show size-dependent blue shift in small noble metal nanowires \cite{Toscano:2015iw}. However, a more detailed comparison with DFT results \cite{Yan:2015ff,Ciraci:2016il} brought out some of the limitations of Toscano's approach and pointed out that in order to describe well both near- and far-field features, one has to sacrifice the self-consistency of the method \cite{Ciraci:2016il}. 
Another unsolved limitation in the QHT is that the size-dependent broadening is completely neglected, although a recent effort in this direction \cite{Li:2015io} proposed the use of a density-dependent damping rate.
Moreover, from a fundamental point of view, a formal derivation of the QHT from first-principles is still missing.

In this letter, I show that the QHT equation can be directly obtained form the single particle Kohn-Sham equation. Moreover, because the QHT intrinsically describes both longitudinal and transverse fields, it is possible to consider a hamiltonian that includes external vector potentials.
This allows to straightforwardly include in the QHT description the current-dependent exchange-correlation (XC) potential developed by Vignale-Kohn \cite{Vignale:1996hk} in the context of time-domain current density functional theory (TD-CDFT).
The results is a self-consistent theory derived form first-principles that can be applied to macroscopic systems.
This theory correctly predicts size-dependent plasmon energies as well as size-dependent broadening.
Moreover, my implementation can be applied to nanoparticle dimers for which results show good agreement in comparison with DFT calculations previously published.

As a first step I will show that the QHT equation can be derived directly from the single particle Kohn-Sham (KS). This derivation is useful to understand the degree of approximations that are made when using the QHT.
Let us consider a system of $N_e$ noninteracting particles in the presence of an electromagnetic field generated by the scalar and vector potentials $v_{\rm e}({\bf{r}},t)$  and ${\bf A}_{\rm m}( {\bf r},t)$, and in the presence of the XC potential energy $v_{\rm xc}({\bf{r}},t)$ and vector potential ${\bf A}_{\rm xc}( {\bf r},t)$. The system is described by a set of time-dependent KS equations for the single orbitals $\varphi_j({\bf r},t)$, $j=1...N_e$:
\begin{equation} 
i\hbar \frac{{\partial {\varphi _j}}}{{\partial t}} = \left[ {\frac{{{{\left( {i\hbar \nabla  - e{\bf{A}}} \right)}^2}}}{{2m}} - e{v_{\rm{e}}} + {v_{{\rm{xc}}}}} \right]{\varphi _j},
\label{eq:KS}
\end{equation} 
\normalsize
where $\hbar$ is the reduced Planck constant, $m$ and $e$ are the electron mass and charge (in absolute value), respectively, and ${\bf{A}}={\bf{A}}_{\rm m}+{\bf{A}}_{\rm xc}$. The electromagnetic potentials are related to the usual electric and magnetic fields $\bf E$ and $\bf B$, respectively, via the relations: ${\bf{E}} =  - \frac{{\partial {\bf{A}}_{\rm m}}}{{\partial t}} - \nabla v_{\rm e}$ and ${\bf{B}} = \nabla  \times {\bf{A}}_{\rm m}$.

Without loss of generality we can write the complex eigenfunctions as $\varphi_j=\phi_j e^{i\chi_j}$ with $\phi_j({\bf r},t)$ and $\chi_j({\bf r},t)$ purely real functions of space and time.
Our goal is to express Eq. (\ref{eq:KS}) as a function of the global macroscopic variables: the particle density $n({\bf r},t)$, and the density of current ${\bf J}({\bf r},t)$, which are defined as:
\small
\begin{equation} 
\begin{array}{*{20}{l}}
{n = \sum\limits_{j \in {\rm{occ}}} {\phi _j^2} },\\
{{\bf{J}} =  - e\sum\limits_{j \in {\rm{occ}}} {\phi _j^2{{\bf{v}}_j}}  - \frac{{n{e^2}}}{m}{\bf{A}}},
\end{array}
\label{eq:nJ}
\end{equation} 
\normalsize
where the sum is performed over all occupied states and the particle velocities have been defined as ${\bf v}_j=\frac{\hbar}{m}\nabla \chi_j$. Substituting Eq. (\ref{eq:nJ}) into Eq. (\ref{eq:KS}), after some tedious but simple algebraic manipulation, we obtain:
\begin{equation} 
\begin{array}{l}
\frac{{\partial {\bf{J}}}}{{\partial t}} = \frac{{n{e^2}}}{m}{\bf{E}} - \frac{e}{m}{\bf{J}} \times \left( {{\bf{B}} + \nabla  \times {{\bf{A}}_{{\rm{xc}}}}} \right)+\\
\quad \quad   + \frac{{n{e}}}{m}\left( {\nabla {v_{{\rm{xc}}}} -e \frac{{\partial {{\bf{A}}_{{\rm{xc}}}}}}{{\partial t}}} \right) + \frac{e}{m}\nabla  \cdot \Pi ,
\end{array}
\label{eq:exactQHT1}
\end{equation} 
where the momentum flux tensor $\Pi$ is given by:
\begin{equation} 
\begin{array}{*{20}{l}}
{{\Pi _{\mu \nu }} = \frac{{{\hbar ^2}}}{{2m}}\left( { - \frac{{{\delta _{\mu \nu }}}}{2}{\nabla ^2}n + 2\sum\limits_{j \in {\rm{occ}}} {\frac{{\partial {\phi _j}}}{{\partial {r_\mu }}}\frac{{\partial {\phi _j}}}{{\partial {r_\nu }}}} } \right) + }\\
{\quad \quad \quad  + m\sum\limits_{j \in {\rm{occ}}} {\phi _j^2\left( {{v_{\mu ,j}} + \frac{e}{m}{A_\mu }} \right)\left( {{v_{\nu ,j}} + \frac{e}{m}{A_\nu }} \right)} },
\end{array}
\label{eq:tensor1}
\end{equation} 
where the subscript $\mu$ and $\nu$ span the cartesian directions $x$, $y$ and $z$, and $\delta_{\mu\nu}$ is the Kronecker delta. 

It is useful to extract from the sums in Eq. (\ref{eq:tensor1}) the known quantities. In order to do so let us write the single orbitals as the difference, $ {{\tilde \phi }_j}$, with respect to an average orbital defined as $\phi  = \sqrt {\frac{n}{N_e}}$, so that we have ${\phi _j} = \phi  + {{\tilde \phi }_j}$.
Analogously, for the velocities we have ${{\bf{v}}_j} = {\bf{v}} + {{{\bf{\tilde v}}}_j}$, where ${\bf v} = {\bf J}/(-e n)-e{\bf A}/m$ (note that by definition $\sum\limits_{j \in {\rm{occ}}} {\phi _j^2{{{\bf{\tilde v}}}_j}}  = 0$).
After using the new definitions, Eq. (\ref{eq:exactQHT1}) takes the form:
\begin{equation} 
\begin{array}{l}
\frac{{\partial {\bf{J}}}}{{\partial t}} = \frac{{n{e^2}}}{m}{\bf{E}} - \frac{e}{m}{\bf{J}} \times \left( {{\bf{B}} + \nabla  \times {{\bf{A}}_{{\rm{xc}}}}} \right) + \frac{{n{e}}}{m}\left( {\nabla {v_{{\rm{xc}}}} -e \frac{{\partial {{\bf{A}}_{{\rm{xc}}}}}}{{\partial t}}} \right) + \\
\quad \quad \quad  + \frac{ne}{m}\nabla \frac{{\delta {T_{\rm{w}}}}}{{\delta n}} + \frac{1}{e}\left( {\frac{{\bf{J}}}{n}\nabla  \cdot {\bf{J}} + {\bf{J}} \cdot \nabla \frac{{\bf{J}}}{n}} \right) + \frac{e}{m}\nabla  \cdot \Pi ',
\end{array}
\label{eq:exactQHT2}
\end{equation}
 \normalsize
where $\frac{\delta T_{\rm{W}}}{\delta n}=\frac{\hbar^2}{8m}(\frac{\nabla n\cdot\nabla n}{n^2}-2\frac{\nabla^2n}{n})$ is the von Weizs\"acker kinetic potential, and the remaining part of the momentum flux tensor is:
\begin{equation} 
\begin{array}{*{20}{l}}
{\Pi {'_{\mu \nu }} = \frac{{{\hbar ^2}}}{m}\sum\limits_{j \in {\rm{occ}}} {\frac{{\partial {{\tilde \phi }_j}}}{{\partial {r_\mu }}}\frac{{\partial \phi }}{{\partial {r_\nu }}} + \frac{{\partial \phi }}{{\partial {r_\mu }}}\frac{{\partial {{\tilde \phi }_j}}}{{\partial {r_\nu }}} + \frac{{\partial {{\tilde \phi }_j}}}{{\partial {r_\mu }}}\frac{{\partial {{\tilde \phi }_j}}}{{\partial {r_\nu }}}}  + }\\
{\quad \quad \quad \quad \quad  + m\sum\limits_{j \in {\rm{occ}}} {\phi _j^2{{\tilde v}_{\mu ,j}}{{\tilde v}_{\nu ,j}}} }.
\end{array}
\label{eq:tensor2}
\end{equation}
It is worth noting that no approximations have been made up to this point. 
In particular, for the simple case of $N_e=2$ (single orbital) it easy to show that $\Pi {'_{\mu \nu }}=0$.
In fact because there is one occupied orbital $\phi_1=\sqrt {n/2}$, the electrons move in phase, hence $\tilde {\bf  v}_1=0$.
Equation (\ref{eq:exactQHT2}) is then an exact hydrodynamic description of the two-electrons system. Solving this equation would give the exact same result of Eq. (\ref{eq:KS}).
In general, however, in its present form (for $\Pi {'_{\mu \nu }}\ne0$), Eq. (\ref{eq:exactQHT2}) cannot be solved without having information on the single orbitals $\phi_j$.

Our next step is then finding an approximation for $\Pi {'_{\mu \nu }}$ that will allow its evaluation without having to solve for the single orbitals.
We anticipate that such approximation is in fact the TF contribution to the kinetic energy. In particular, we will show that $\nabla \cdot \Pi' \simeq n\nabla \frac{\delta T_{\rm TF}}{\delta n}$, where $T_{\rm TF}=c_{\rm TF}n^{5/3}$ with $c_{\rm TF}=\frac{\hbar^2}{m}\frac{3}{10}(3\pi)^{2/3}$.

Since we are interested in describing structures that are constituted by a large number of electrons, in first approximation we can consider the electronic system as a homogenous electron gas whose orbitals are $\varphi_{\bf k}= \frac{{{e^{i{\bf{k}} \cdot {\bf{r}}}}}}{{{V^{1/2}}}}$, with $V$ the occupied volume in real space.
It is easy to identify $\hbar{\bf k}=m \tilde {\bf v}_j$ (the velocities must be considered without their net contribution induced by the external fields), so that the sum over $j$ becomes a sum of $\bf k$:
\small
\begin{equation}
\frac{{\partial \Pi {'_{\mu \nu }}}}{{\partial {r_\nu }}} \simeq  m\frac{\partial }{{\partial {r_\nu }}}\sum\limits_{j \in {\rm{occ}}} {\phi _j^2{{\tilde v}_{\mu ,j}}{{\tilde v}_{\nu ,j}}}  \simeq  \frac{{2{\hbar ^2}}}{{mV}}\frac{\partial }{{\partial {r_\nu }}}\sum\limits_{{\bf{k}} \in {\rm{occ}}} {{k_\mu }{k_\nu }},
\end{equation}
\normalsize
where the first sum in Eq. (\ref{eq:tensor2}) is zero. Note that even if every term seems to be constant taking the divergence of the second sum will not give equally zero, because, as it will be clear later, the number of occupied states will depend on the local density $n({\bf r})$. 
Since there are many occupied states, we can replace as usual the sum by an integral:
\begin{equation} 
\begin{array}{l}
 \frac{{2{\hbar ^2}}}{{{m}V}}\frac{\partial }{{\partial {r_\nu }}}\sum\limits_{{\bf{k}} \in {\rm{occ}}} {{k_\mu }{k_\nu }} \\
\quad \quad  \simeq \frac{{ {\hbar ^2}}}{{{m}4{\pi ^3}}}\frac{\partial }{{\partial {r_\nu }}}\int\limits_0^{{k_{\rm{F}}}} {{k^4}dk} \int\limits_0^\pi  {\int\limits_0^{2\pi } {{{\hat k}_\mu }{{\hat k}_\nu }} } \sin \theta d\theta d\phi 
\end{array}
\end{equation}
where we used $dn=\frac{V}{8\pi^3}d{\bf k}$ and the Fermi wavenumber is $k_{\rm F}({\bf r})=(3\pi^2n)^{1/3}$. Evaluating the integrals gives: 
\begin{equation} 
 \nabla  \cdot \Pi ' \simeq \frac{{ {\hbar ^2}}}{{{m}{\pi ^2}}}\frac{1}{{15}}\frac{\partial }{{\partial {r_\nu }}}\left( {{\delta _{\mu \nu }}k_{\rm{F}}^5} \right) = \frac{{10}}{9}{c_{{\rm{TF}}}}{n^{2/3}}\nabla n
 \label{eq:TF}
\end{equation}
 \normalsize
It is easy now to show that the last term in the previous equation is exactly equal to $n\nabla \frac{\delta T_{\rm TF}}{\delta n}$, as it was anticipated.
The previous derivation demonstrates that the error committed when using the QHT is solely given by the approximation in Eq. (\ref{eq:TF}).

It remains now to give an explicit expression for the XC potentials. For the scalar potential $v_{\rm xc}$ the usual local density approximation (LDA) is assumed \cite{Perdew:1981dv}.
Here, however, we would like to go beyond the LDA in order to provide a damping mechanism for collective excitations.
In fact, while in DFT single particles can be excited into particle-hole pairs, providing then a form of broadening of the collective resonances, in QHT all the particles are assumed to lay in identical states.
The XC potential however is an intrinsically non-local functional of the density, namely, it does not admit a gradient expansion in $n$ without sacrificing some basic symmetries \cite{Vignale:1996hk,Vignale:1997jda}.
Fortunately, Vignale and Kohn have shown in the context of TD-CDFT that a local gradient expansion is still possible in terms of the current density $\bf J$, and have provided an explicit approximated expression for the XC vector potential ${\bf A}_{\rm xc}$ \cite{Vignale:1996hk}.
It was later shown that the XC vector potential can be arranged into a more intuitive form \cite{Vignale:1997jda}, so that it is expressed as the divergence of the viscoelastic stress tensor, namely: 
\begin{equation}
\frac{{\partial {{\bf{A}}_{{\rm{xc}}}}}}{{\partial t}} = \frac{1}{en}\nabla  \cdot  \sigma,
\label{eq:Axc}
\end{equation}
\normalsize
where $\sigma$ is the classical viscoelastic stress tensor:
\small
\begin{equation}
{\sigma _{\mu \nu }} = \tilde \eta \left( {\frac{{\partial {v_\mu }}}{{\partial {r_\nu }}} + \frac{{\partial {v_\nu }}}{{\partial {r_\mu }}} - \frac{2}{3}{\delta _{\mu \nu }}\nabla  \cdot {\bf{v}}} \right) + \tilde \zeta {\delta _{\mu \nu }}\nabla  \cdot {\bf{v}},
\label{eq:vet}
\end{equation}
\normalsize
with $\tilde \eta$ and $\tilde \zeta$ generalized complex viscosities that depend on the density $n$ and the frequency $\omega$, and can related to the $k\to0$ limit of the XC longitudinal and transverse kernel functions $f_{\rm xc,L(T)}(\omega,{\bf{k}})$ \cite{Vignale:1997jda}.
The caveat is that the kernel functions are not known exactly, although several interpolation formulae have been developed \cite{Gross:1985gx,Iwamoto:1987es,Nifosi:1998fa,Qian:2002dl}.
In this work the interpolation proposed by Conti and Vignale (CV) is used \cite{Conti:1999fv}, although a modification is introduced in order to take into account the effective broadening of the plasmonic resonance.
Let us write the complex viscosities as:
\begin{equation}
\begin{array}{l}
\tilde \eta \left( {\omega ,{\bf{r}}} \right) = \eta \left( {\bf{r}} \right) - \frac{{\mu \left( {\bf{r}} \right)}}{{i\omega }},\\
\tilde \zeta \left( {\omega ,{\bf{r}}} \right) = \zeta \left( {\bf{r}} \right) -\frac{{K\left( {\bf{r}} \right)}}{{i\omega }}, 
\end{array}
\end{equation}
where the coefficients $\eta$, $\zeta$, $\mu$ and $K$ are the shear and the bulk viscosity, and the shear and the bulk modulus respectively. In the CV approximation all these coefficients are real quantities independent of the frequency, whose interpolation formulas are reported in \footnote{\textit{Supplemental Material}}.
While $\mu$ and $K$ are retrieved so that the limit for low densities is respected (this is important for the study of surface effects in finite structures because the electron density drops to zero in the region of interest), this is not true for $\eta$ for which the low-density limit is unknown ($\zeta$ is identically zero).
We found that best results are obtained if $\eta=15\eta_{\rm CV}$. 

The nonlinear hydrodynamic Eq. (\ref{eq:exactQHT2}) with the approximation in Eq. (\ref{eq:TF}) and the expression in Eq. (\ref{eq:Axc}) constitutes the main result of this letter.
This equation can be coupled to Maxwell's equations to describe self-consistently the linear and nonlinear electromagnetic response of a free-electron gas. 
Having derived our equation from first principles clarifies the level of approximations that are made when employing the QHT in describing the electron dynamic. 
Although, the equation obtained is very similar to the QHT equation used in previous works \cite{Yan:2015ff,Ciraci:2016il,Toscano:2015iw,Li:2015io} (except for the introduction of the XC vector potential ${\bf A}_{\rm xc}$), the present derivation allows to better define some critical terms.
The von  Weizs\"acker term for example is usually preceded by a fraction that might vary from $1/9$ to $1$.
It is clear from Eq. (\ref{eq:exactQHT2}) that the right choice should be $1$ as already suggested in Refs.  \cite{Yan:2015ff,Ciraci:2016il} by direct comparisons with DFT results.

Moreover, the fact that the viscoelastic coefficients in Eq. (\ref{eq:vet}) are complex quantities adds to the theory two important improvements: i) the XC viscosity provide a $k$-dependent damping mechanism for collective excitations; and ii) it introduces a correction to the plasmon dispersion, which takes into account the fact that the local Fermi surface deviates from its quasi-equilibrium shape \cite{Giuliani:2005ve}.

As a first application of Eq. (\ref{eq:exactQHT2}), let us consider the linear optical response of single jellium nanospheres. 
By linearizing Eq. (\ref{eq:exactQHT2}), coupling it to Maxwell's equations, and remembering that  ${\partial{\bf P}}/{\partial t}={\bf J}$, we obtain in the frequency domain the following system of equations:
\begin{equation} 
\begin{array}{l}
\nabla  \times \nabla  \times {\bf{E}} - \frac{{{\omega ^2}}}{{{c^2}}}{\bf{E}} = {\omega ^2}{\mu _0}{\bf{P}},\\
 - \frac{{e{n_0}}}{m}\nabla {\left( {\frac{{\delta G}}{{\delta n}}} \right)_1} + \frac{i\omega}{m}\nabla  \cdot \sigma[\frac{\bf P}{n_0}]  - \left( {{\omega ^2} + i\omega \gamma } \right){\bf{P}} = \frac{{{e^2}{n_0}}}{m}{\bf{E}},
\end{array}
\label{eq:sys}
\end{equation}
where $n_0$ is the equilibrium density, $\mu_0$ and $c$ are the magnetic permeability and the speed of light in vacuum, respectively; $\frac{\delta G}{\delta n}=\frac{\delta T_{\rm TF}}{\delta n}+\frac{\delta T_{\rm W}}{\delta n}+v_{\rm xc}$ (explicits expressions for  ${\left( {\frac{{\delta G}}{{\delta n}}} \right)}_1$ can be found in Ref. \cite{Ciraci:2016il}).
In writing the second Eq. (\ref{eq:sys}) the phenomenological damping rate $\gamma$ has been introduced in order to take into account losses occurring in the bulk region.
The ground state $n_0$ can be calculated self-consistently \cite{Note1} by solving the following differential equation \cite{Toscano:2015iw}:
\small
\begin{equation} 
{\nabla ^2}{\left( {\frac{{\delta G[n]}}{{\delta n}}} \right)_{n=n_0}}  + \frac{{{e^2}}}{{{\epsilon _0}}}\left( { {n_0}-{n^ + }} \right) = 0,
\label{eq:n0}
\end{equation}
\normalsize
where $\epsilon_0$ is the electric permittivity and $n^+$ is the homogenous ion density.
Note that since ${\bf A}_{\rm xc}$ affects only the dynamical response, Eq. (\ref{eq:n0}) and the properties of its solution $n_0$ remain unchanged with respect to Ref. \cite{Ciraci:2016il}. 
 
The system of Eqs. (\ref{eq:sys}) and (\ref{eq:n0}) is numerically solved with a commercially available software based on the finite-element method, \textsc{Comsol} Multiphysics \footnote{\textsc{Comsol} Multiphysics, http://www.comsol.com}.
In particular, the 2.5D technique \cite{Ciraci:2013wi} has been used, which allows to efficiently compute absorption spectra for axis symmetric structures (see \cite{Note1}).
\begin{figure}
  \centering
 \includegraphics[width=0.48\textwidth]{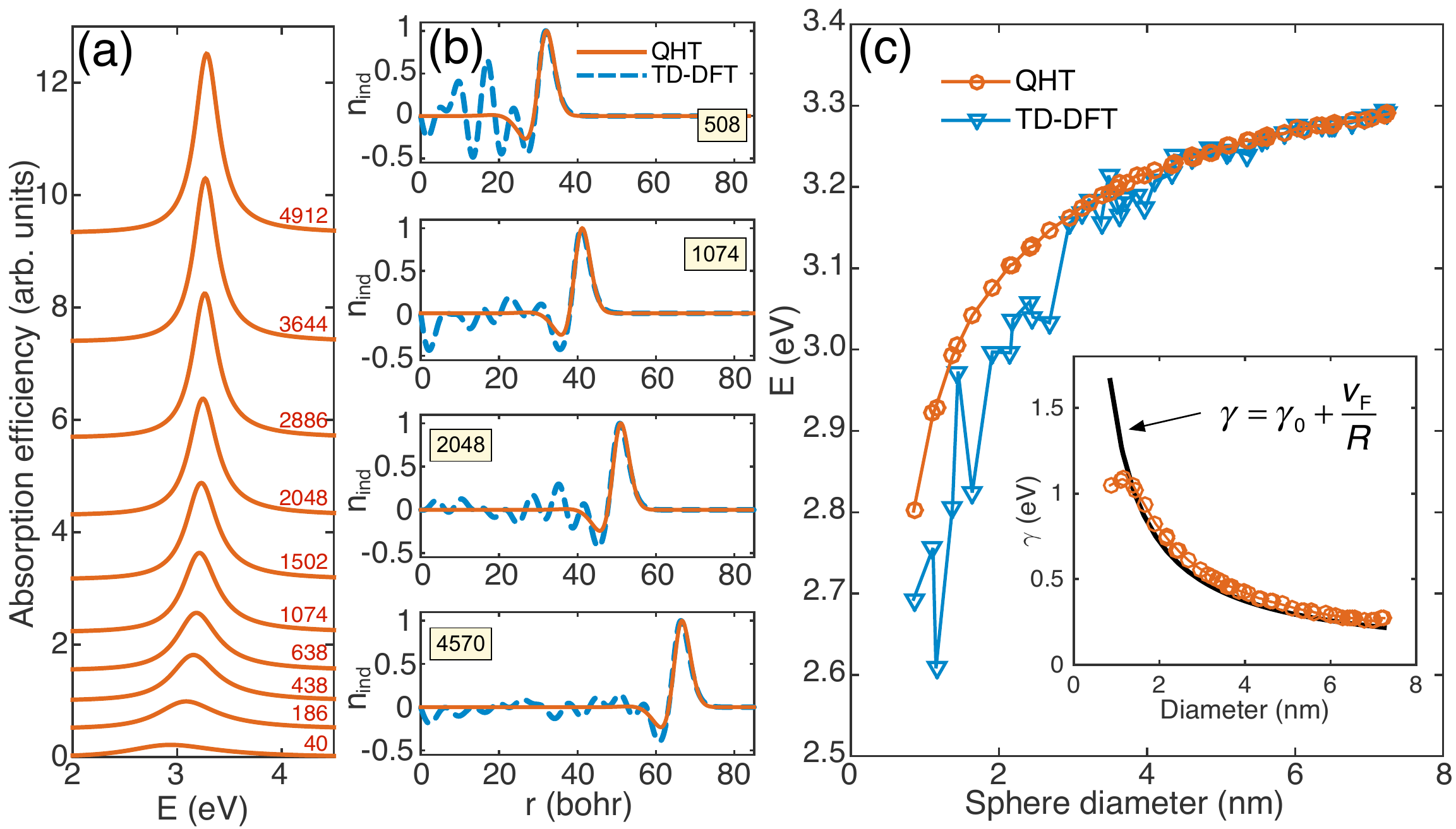}
 \caption{\label{spheres} Properties of jellium spheres ($r_s = 4$ a.u.) of different sizes;  (a) absorption cross-section.  (b) imaginary part of the normalized induced charge density; (c) plasmon resonance as a function of the sphere diameter; in the inset, the broadening of the resonance for QHT. Peak positions and widths were calculated by fitting the spectra with a Lorentzian shaped function; TD-DFT data are taken from Ref. \cite{Ciraci:2016il}.}
 \end{figure}
 Absorption spectra for different jellium Na ($r_s=4$ a.u.) nanospheres are shown in Fig. \ref{spheres}a.
The first thing to notice is that as the particle size shrinks the broader their spectra.
One important difference with previous QHT results\cite{Ciraci:2016il} is the absence of higher energy resonances.
These resonances are the analogue of Rydberg states for atoms. They are associated with very delocalized states and are numerically affected by the finiteness of the simulation domain size\cite{Ciraci:2016il}.
With the introduction of the XC viscosity, these states are no longer supported as it would be expected for jellium spheres.

In Fig. \ref{spheres}b the QHT induced charge densities (imaginary part) corresponding to plasmonic peak resonances is compared to the TD-DFT results.
Although oscillations appearing in the TD-DFT case in the bulk region are not reproduced, the main induced peak is very well described by the QHT.
Remark that this is not necessarily the case for the approach used in Ref. \cite{Li:2015io} where a damping factor $\gamma \propto n_0^{-5/6}$ is assumed.
Because $\gamma$ diverges when $n_0$ goes to zero, \textit{i.e.} near the particle surface, the induced density is \textit{prematurely} damped at the surface. 

In Fig. \ref{spheres}c the plasmon resonances obtained with the present QHT model is compared against TD-DFT results\cite{Ciraci:2016il}, for nanoparticle diameters $D$ ranging from $\sim0.85$ to $\sim7.25$ nm ($N_e=8$ to $N_e=5032$).
For $D>3$ nm ($N_e > 398$) QHT reproduces DFT plasmon energies with great accuracy, with QHT resonances marking almost exactly the mean trajectory of DFT data.
Also striking is the comparison of the broadening of the resonance shown in the inset.
The reference curve in this case is given by the known formula \cite{Kreibig:2013ci} $\gamma=\gamma_0+{v_{\rm F}}/{R}$ where $v_{\rm F}$ is the Fermi velocity for the homogeneous electron gas and $R=D/2$.
The agreement is perfect for all the diameters except the smallest ones for which however the analytical formula is not expected to hold. 

\begin{figure}
  \centering
 \includegraphics[width=0.48\textwidth]{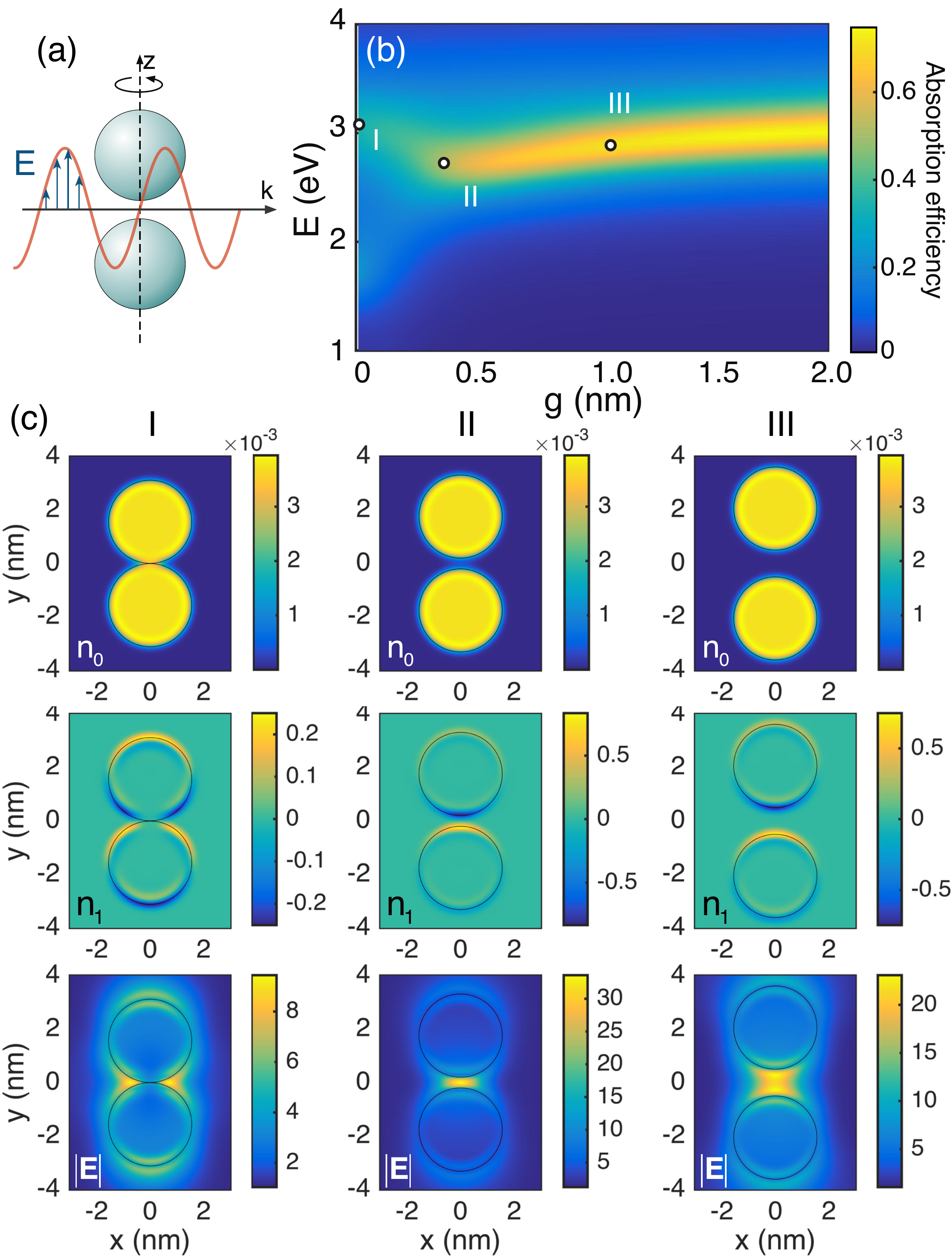}
 \caption{\label{dimer} Dimer of Na spheres constituted by $N_e = 398$ electrons each. (a) Scheme of the system; (b) absorption efficiency spectra as a function of the particle distance $g$; (c) near-field properties corresponding to the points depicted in (b) ($g=1.0$, $0.4$, $0$ nm). The densities $n_0$ and $n_1$ are in atomic units , while $|{\bf E}|$ is normalized to the incident filed amplitude.}
 \end{figure}

Another important system that is worth benchmarking the present QHT model on is the nanoparticle dimer.
As the distance between two closely spaced nanoparticles reduces, four different effects come simultaneously into play \cite{Zuloaga:2009gm,Ciraci:2012fp,Raza:2011io,Savage:2012by}: i) the resonance shift due to the hybridization of the plasmonic modes; ii) effects due to the nonlocal optical response of the electron gas; iii) the broadening of the resonance, which is intrinsically due to the nonlocal absorption (since the size of the spheres remains unchanged); iv) tunneling effects due to the overlap of the electron densities of each particle.
Let us consider a dimer of Na spheres of diameter $D\simeq 3$ nm  ($N_e = 398$) and separated by a distance $g$ that goes from  2 to 0 nm.
The dimer is excited by a plane wave propagating orthogonally to the dimer axis whose electric field is polarized along $z$, as depicted in Fig. \ref{dimer}a. 
Note that we compute the ground state charge density self-consistently using Eq. (\ref{eq:n0}) for each value of the distance $g$. 
The map of Fig. \ref{dimer}b shows the absorption spectrum of the dimer as a function of the gap size.
As the gap shrinks the plasmon resonance undergoes a redshift up to the point ($g\simeq0.4$ nm) where tunneling effects kick in and the resonance broadens and the shift pushes back to higher energies.
Note that the QHT without the XC viscosity would have predicted an unnoticeable broadening.

In Fig. \ref{dimer}c are reported the equilibrium charge density $n_0$, the induced electron density $n_1$ and the electric field norm distribution, respectively, for three critical situations labeled in the spectrum map. 
These results can be directly compared to results of Ref. \cite{Barbry:2015iw} in which time-domain DFT calculations for the same jellium Na dimer are reported. 
It can be seen that all quantities are accurately reproduced.
It is worth noting that in DFT there is no intrinsic broadening mechanism for each spectral line and a phenomenological value of $\gamma$ (usually much larger than the bulk value) has to be taken into account in order to produce a continuous spectrum.

I have presented a QHT model obtained from first principles, which is able to accurately and self-consistently describe far-field and near-field properties of plasmonic systems in most extreme conditions (\textit{i.e.}, tunneling regime).
This model is computationally efficient and can be applied to mesoscopic structures in which quantum effects might be important.
Moreover, it represents a general theory that is also valid in the nonlinear regime\cite{Vignale:1997jda} and can be used for the investigation of optical nonlinear surface effects.

Finally, in the limit of a uniform electron density, the polarization equation in (\ref{eq:sys}) reduces to the generalized non-local optical response model presented in Ref. \cite{Mortensen:2014kc}, with the only difference that the divergence of the viscoelastic stress tensor gives an extra contributions proportional to $\nabla^2 {\bf J}$, other than the one of the form $\nabla\nabla\cdot {\bf J}$.
Although the theory showed in Ref.  \cite{Mortensen:2014kc} seems to work fairly well, its derivation is somewhat questionable and unsatisfactory.
The present QHT gives a proper basis on the model and generalizes it by adding an extra term.

I believe that this model offers a valid and computationally efficient solution for studying the electron dynamic of large plasmonic systems.
It gives access to an unparalleled regime of light-matter interactions, which in turn might lead to novel and unexploited effects.\\

\begin{acknowledgements}
The author thanks Dr. Fabio Della Sala for fruitful discussions.
\end{acknowledgements}


\begin{thebibliography}{40}%
\makeatletter
\providecommand \@ifxundefined [1]{%
 \@ifx{#1\undefined}
}%
\providecommand \@ifnum [1]{%
 \ifnum #1\expandafter \@firstoftwo
 \else \expandafter \@secondoftwo
 \fi
}%
\providecommand \@ifx [1]{%
 \ifx #1\expandafter \@firstoftwo
 \else \expandafter \@secondoftwo
 \fi
}%
\providecommand \natexlab [1]{#1}%
\providecommand \enquote  [1]{``#1''}%
\providecommand \bibnamefont  [1]{#1}%
\providecommand \bibfnamefont [1]{#1}%
\providecommand \citenamefont [1]{#1}%
\providecommand \href@noop [0]{\@secondoftwo}%
\providecommand \href [0]{\begingroup \@sanitize@url \@href}%
\providecommand \@href[1]{\@@startlink{#1}\@@href}%
\providecommand \@@href[1]{\endgroup#1\@@endlink}%
\providecommand \@sanitize@url [0]{\catcode `\\12\catcode `\$12\catcode
  `\&12\catcode `\#12\catcode `\^12\catcode `\_12\catcode `\%12\relax}%
\providecommand \@@startlink[1]{}%
\providecommand \@@endlink[0]{}%
\providecommand \url  [0]{\begingroup\@sanitize@url \@url }%
\providecommand \@url [1]{\endgroup\@href {#1}{\urlprefix }}%
\providecommand \urlprefix  [0]{URL }%
\providecommand \Eprint [0]{\href }%
\providecommand \doibase [0]{http://dx.doi.org/}%
\providecommand \selectlanguage [0]{\@gobble}%
\providecommand \bibinfo  [0]{\@secondoftwo}%
\providecommand \bibfield  [0]{\@secondoftwo}%
\providecommand \translation [1]{[#1]}%
\providecommand \BibitemOpen [0]{}%
\providecommand \bibitemStop [0]{}%
\providecommand \bibitemNoStop [0]{.\EOS\space}%
\providecommand \EOS [0]{\spacefactor3000\relax}%
\providecommand \BibitemShut  [1]{\csname bibitem#1\endcsname}%
\let\auto@bib@innerbib\@empty
\bibitem [{\citenamefont {Raza}\ \emph {et~al.}(2011)\citenamefont {Raza},
  \citenamefont {Toscano}, \citenamefont {Jauho}, \citenamefont {Wubs},\ and\
  \citenamefont {Mortensen}}]{Raza:2011io}%
  \BibitemOpen
  \bibfield  {author} {\bibinfo {author} {\bibfnamefont {S.}~\bibnamefont
  {Raza}}, \bibinfo {author} {\bibfnamefont {G.}~\bibnamefont {Toscano}},
  \bibinfo {author} {\bibfnamefont {A.~P.}\ \bibnamefont {Jauho}}, \bibinfo
  {author} {\bibfnamefont {M.}~\bibnamefont {Wubs}}, \ and\ \bibinfo {author}
  {\bibfnamefont {N.~A.}\ \bibnamefont {Mortensen}},\ }\href@noop {} {\bibfield
   {journal} {\bibinfo  {journal} {Phys. Rev. B}\ }\textbf {\bibinfo {volume}
  {84}},\ \bibinfo {pages} {121412} (\bibinfo {year} {2011})}\BibitemShut
  {NoStop}%
\bibitem [{\citenamefont {Fern{\'a}ndez-Dom{\'\i}nguez}\ \emph
  {et~al.}(2012)\citenamefont {Fern{\'a}ndez-Dom{\'\i}nguez}, \citenamefont
  {Zhang}, \citenamefont {Luo}, \citenamefont {Maier}, \citenamefont
  {Garc{\'\i}a-Vidal},\ and\ \citenamefont
  {Pendry}}]{FernandezDominguez:2012eg}%
  \BibitemOpen
  \bibfield  {author} {\bibinfo {author} {\bibfnamefont {A.~I.}\ \bibnamefont
  {Fern{\'a}ndez-Dom{\'\i}nguez}}, \bibinfo {author} {\bibfnamefont
  {P.}~\bibnamefont {Zhang}}, \bibinfo {author} {\bibfnamefont
  {Y.}~\bibnamefont {Luo}}, \bibinfo {author} {\bibfnamefont {S.~A.}\
  \bibnamefont {Maier}}, \bibinfo {author} {\bibfnamefont {F.~J.}\ \bibnamefont
  {Garc{\'\i}a-Vidal}}, \ and\ \bibinfo {author} {\bibfnamefont {J.~B.}\
  \bibnamefont {Pendry}},\ }\href@noop {} {\bibfield  {journal} {\bibinfo
  {journal} {Phys. Rev. B}\ }\textbf {\bibinfo {volume} {86}},\ \bibinfo
  {pages} {241110} (\bibinfo {year} {2012})}\BibitemShut {NoStop}%
\bibitem [{\citenamefont {Toscano}\ \emph {et~al.}(2012)\citenamefont
  {Toscano}, \citenamefont {Raza}, \citenamefont {Jauho}, \citenamefont
  {Mortensen},\ and\ \citenamefont {Wubs}}]{Toscano:2012fh}%
  \BibitemOpen
  \bibfield  {author} {\bibinfo {author} {\bibfnamefont {G.}~\bibnamefont
  {Toscano}}, \bibinfo {author} {\bibfnamefont {S.}~\bibnamefont {Raza}},
  \bibinfo {author} {\bibfnamefont {A.-P.}\ \bibnamefont {Jauho}}, \bibinfo
  {author} {\bibfnamefont {N.~A.}\ \bibnamefont {Mortensen}}, \ and\ \bibinfo
  {author} {\bibfnamefont {M.}~\bibnamefont {Wubs}},\ }\href@noop {} {\bibfield
   {journal} {\bibinfo  {journal} {Opt. Express}\ }\textbf {\bibinfo {volume}
  {20}},\ \bibinfo {pages} {4176} (\bibinfo {year} {2012})}\BibitemShut
  {NoStop}%
\bibitem [{\citenamefont {Cirac{\`\i}}\ \emph
  {et~al.}(2013{\natexlab{a}})\citenamefont {Cirac{\`\i}}, \citenamefont
  {Urzhumov},\ and\ \citenamefont {Smith}}]{Ciraci:2013jt}%
  \BibitemOpen
  \bibfield  {author} {\bibinfo {author} {\bibfnamefont {C.}~\bibnamefont
  {Cirac{\`\i}}}, \bibinfo {author} {\bibfnamefont {Y.~A.}\ \bibnamefont
  {Urzhumov}}, \ and\ \bibinfo {author} {\bibfnamefont {D.~R.}\ \bibnamefont
  {Smith}},\ }\href@noop {} {\bibfield  {journal} {\bibinfo  {journal} {J. Opt.
  Soc. Am. B}\ }\textbf {\bibinfo {volume} {30}},\ \bibinfo {pages} {2731}
  (\bibinfo {year} {2013}{\natexlab{a}})}\BibitemShut {NoStop}%
\bibitem [{\citenamefont {Yan}\ and\ \citenamefont
  {Mortensen}(2016)}]{Yan:2016gz}%
  \BibitemOpen
  \bibfield  {author} {\bibinfo {author} {\bibfnamefont {W.}~\bibnamefont
  {Yan}}\ and\ \bibinfo {author} {\bibfnamefont {N.~A.}\ \bibnamefont
  {Mortensen}},\ }\href@noop {} {\bibfield  {journal} {\bibinfo  {journal}
  {Phys. Rev. B}\ }\textbf {\bibinfo {volume} {93}},\ \bibinfo {pages} {115439}
  (\bibinfo {year} {2016})}\BibitemShut {NoStop}%
\bibitem [{\citenamefont {Zuloaga}\ \emph {et~al.}(2009)\citenamefont
  {Zuloaga}, \citenamefont {Prodan},\ and\ \citenamefont
  {Nordlander}}]{Zuloaga:2009gm}%
  \BibitemOpen
  \bibfield  {author} {\bibinfo {author} {\bibfnamefont {J.}~\bibnamefont
  {Zuloaga}}, \bibinfo {author} {\bibfnamefont {E.}~\bibnamefont {Prodan}}, \
  and\ \bibinfo {author} {\bibfnamefont {P.}~\bibnamefont {Nordlander}},\
  }\href@noop {} {\bibfield  {journal} {\bibinfo  {journal} {Nano Lett.}\
  }\textbf {\bibinfo {volume} {9}},\ \bibinfo {pages} {887} (\bibinfo {year}
  {2009})}\BibitemShut {NoStop}%
\bibitem [{\citenamefont {Zhu}\ \emph {et~al.}(2016)\citenamefont {Zhu},
  \citenamefont {Esteban}, \citenamefont {Borisov}, \citenamefont {Baumberg},
  \citenamefont {Nordlander}, \citenamefont {Lezec}, \citenamefont {Aizpurua},\
  and\ \citenamefont {Crozier}}]{Zhu:2016fn}%
  \BibitemOpen
  \bibfield  {author} {\bibinfo {author} {\bibfnamefont {W.}~\bibnamefont
  {Zhu}}, \bibinfo {author} {\bibfnamefont {R.}~\bibnamefont {Esteban}},
  \bibinfo {author} {\bibfnamefont {A.~G.}\ \bibnamefont {Borisov}}, \bibinfo
  {author} {\bibfnamefont {J.~J.}\ \bibnamefont {Baumberg}}, \bibinfo {author}
  {\bibfnamefont {P.}~\bibnamefont {Nordlander}}, \bibinfo {author}
  {\bibfnamefont {H.~J.}\ \bibnamefont {Lezec}}, \bibinfo {author}
  {\bibfnamefont {J.}~\bibnamefont {Aizpurua}}, \ and\ \bibinfo {author}
  {\bibfnamefont {K.~B.}\ \bibnamefont {Crozier}},\ }\href@noop {} {\bibfield
  {journal} {\bibinfo  {journal} {Nat. Comm.}\ }\textbf {\bibinfo {volume}
  {7}},\ \bibinfo {pages} {1} (\bibinfo {year} {2016})}\BibitemShut {NoStop}%
\bibitem [{\citenamefont {Teperik}\ \emph {et~al.}(2013)\citenamefont
  {Teperik}, \citenamefont {Nordlander}, \citenamefont {Aizpurua},\ and\
  \citenamefont {Borisov}}]{Teperik:2013dd}%
  \BibitemOpen
  \bibfield  {author} {\bibinfo {author} {\bibfnamefont {T.~V.}\ \bibnamefont
  {Teperik}}, \bibinfo {author} {\bibfnamefont {P.}~\bibnamefont {Nordlander}},
  \bibinfo {author} {\bibfnamefont {J.}~\bibnamefont {Aizpurua}}, \ and\
  \bibinfo {author} {\bibfnamefont {A.~G.}\ \bibnamefont {Borisov}},\
  }\href@noop {} {\bibfield  {journal} {\bibinfo  {journal} {Phys. Rev. Lett.}\
  }\textbf {\bibinfo {volume} {110}},\ \bibinfo {pages} {263901} (\bibinfo
  {year} {2013})}\BibitemShut {NoStop}%
\bibitem [{\citenamefont {Cirac{\`\i}}\ \emph {et~al.}(2012)\citenamefont
  {Cirac{\`\i}}, \citenamefont {Hill}, \citenamefont {Mock}, \citenamefont
  {Urzhumov}, \citenamefont {Fernandez-Dominguez}, \citenamefont {Maier},
  \citenamefont {J~B}, \citenamefont {Chilkoti},\ and\ \citenamefont
  {Smith}}]{Ciraci:2012fp}%
  \BibitemOpen
  \bibfield  {author} {\bibinfo {author} {\bibfnamefont {C.}~\bibnamefont
  {Cirac{\`\i}}}, \bibinfo {author} {\bibfnamefont {R.~T.}\ \bibnamefont
  {Hill}}, \bibinfo {author} {\bibfnamefont {J.~J.}\ \bibnamefont {Mock}},
  \bibinfo {author} {\bibfnamefont {Y.~A.}\ \bibnamefont {Urzhumov}}, \bibinfo
  {author} {\bibfnamefont {A.~I.}\ \bibnamefont {Fernandez-Dominguez}},
  \bibinfo {author} {\bibfnamefont {S.~A.}\ \bibnamefont {Maier}}, \bibinfo
  {author} {\bibfnamefont {P.}~\bibnamefont {J~B}}, \bibinfo {author}
  {\bibfnamefont {A.}~\bibnamefont {Chilkoti}}, \ and\ \bibinfo {author}
  {\bibfnamefont {D.~R.}\ \bibnamefont {Smith}},\ }\href@noop {} {\bibfield
  {journal} {\bibinfo  {journal} {Science}\ }\textbf {\bibinfo {volume}
  {337}},\ \bibinfo {pages} {1072} (\bibinfo {year} {2012})}\BibitemShut
  {NoStop}%
\bibitem [{\citenamefont {Savage}\ \emph {et~al.}(2012)\citenamefont {Savage},
  \citenamefont {Hawkeye}, \citenamefont {Esteban},\ and\ \citenamefont
  {Borisov}}]{Savage:2012by}%
  \BibitemOpen
  \bibfield  {author} {\bibinfo {author} {\bibfnamefont {K.~J.}\ \bibnamefont
  {Savage}}, \bibinfo {author} {\bibfnamefont {M.~M.}\ \bibnamefont {Hawkeye}},
  \bibinfo {author} {\bibfnamefont {R.}~\bibnamefont {Esteban}}, \ and\
  \bibinfo {author} {\bibfnamefont {A.~G.}\ \bibnamefont {Borisov}},\
  }\href@noop {} {\bibfield  {journal} {\bibinfo  {journal} {Nature}\ }\textbf
  {\bibinfo {volume} {491}},\ \bibinfo {pages} {574} (\bibinfo {year}
  {2012})}\BibitemShut {NoStop}%
\bibitem [{\citenamefont {Scholl}\ \emph {et~al.}(2013)\citenamefont {Scholl},
  \citenamefont {Garc{\'\i}a-Etxarri}, \citenamefont {Koh},\ and\ \citenamefont
  {Dionne}}]{Scholl:2013ge}%
  \BibitemOpen
  \bibfield  {author} {\bibinfo {author} {\bibfnamefont {J.~A.}\ \bibnamefont
  {Scholl}}, \bibinfo {author} {\bibfnamefont {A.}~\bibnamefont
  {Garc{\'\i}a-Etxarri}}, \bibinfo {author} {\bibfnamefont {A.~L.}\
  \bibnamefont {Koh}}, \ and\ \bibinfo {author} {\bibfnamefont {J.~A.}\
  \bibnamefont {Dionne}},\ }\href@noop {} {\bibfield  {journal} {\bibinfo
  {journal} {Nano Lett.}\ }\textbf {\bibinfo {volume} {13}},\ \bibinfo {pages}
  {564} (\bibinfo {year} {2013})}\BibitemShut {NoStop}%
\bibitem [{\citenamefont {Cirac{\`\i}}\ \emph {et~al.}(2014)\citenamefont
  {Cirac{\`\i}}, \citenamefont {Chen}, \citenamefont {Mock}, \citenamefont
  {McGuire}, \citenamefont {Liu}, \citenamefont {Oh},\ and\ \citenamefont
  {Smith}}]{Ciraci:2014jv}%
  \BibitemOpen
  \bibfield  {author} {\bibinfo {author} {\bibfnamefont {C.}~\bibnamefont
  {Cirac{\`\i}}}, \bibinfo {author} {\bibfnamefont {X.}~\bibnamefont {Chen}},
  \bibinfo {author} {\bibfnamefont {J.~J.}\ \bibnamefont {Mock}}, \bibinfo
  {author} {\bibfnamefont {F.}~\bibnamefont {McGuire}}, \bibinfo {author}
  {\bibfnamefont {X.}~\bibnamefont {Liu}}, \bibinfo {author} {\bibfnamefont
  {S.-H.}\ \bibnamefont {Oh}}, \ and\ \bibinfo {author} {\bibfnamefont {D.~R.}\
  \bibnamefont {Smith}},\ }\href@noop {} {\bibfield  {journal} {\bibinfo
  {journal} {Appl. Phys. Lett.}\ }\textbf {\bibinfo {volume} {104}},\ \bibinfo
  {pages} {023109} (\bibinfo {year} {2014})}\BibitemShut {NoStop}%
\bibitem [{\citenamefont {Hajisalem}\ \emph {et~al.}(2014)\citenamefont
  {Hajisalem}, \citenamefont {Nezami},\ and\ \citenamefont
  {Gordon}}]{Hajisalem:2014cr}%
  \BibitemOpen
  \bibfield  {author} {\bibinfo {author} {\bibfnamefont {G.}~\bibnamefont
  {Hajisalem}}, \bibinfo {author} {\bibfnamefont {M.~S.}\ \bibnamefont
  {Nezami}}, \ and\ \bibinfo {author} {\bibfnamefont {R.}~\bibnamefont
  {Gordon}},\ }\href@noop {} {\bibfield  {journal} {\bibinfo  {journal} {Nano
  Lett.}\ }\textbf {\bibinfo {volume} {14}},\ \bibinfo {pages} {6651} (\bibinfo
  {year} {2014})}\BibitemShut {NoStop}%
\bibitem [{\citenamefont {Lin}\ \emph {et~al.}(2015)\citenamefont {Lin},
  \citenamefont {Zapata}, \citenamefont {Xiong}, \citenamefont {Liu},
  \citenamefont {Wang}, \citenamefont {Xu}, \citenamefont {Borisov},
  \citenamefont {Gu}, \citenamefont {Nordlander}, \citenamefont {Aizpurua},\
  and\ \citenamefont {Ye}}]{Lin:2015et}%
  \BibitemOpen
  \bibfield  {author} {\bibinfo {author} {\bibfnamefont {L.}~\bibnamefont
  {Lin}}, \bibinfo {author} {\bibfnamefont {M.}~\bibnamefont {Zapata}},
  \bibinfo {author} {\bibfnamefont {M.}~\bibnamefont {Xiong}}, \bibinfo
  {author} {\bibfnamefont {Z.}~\bibnamefont {Liu}}, \bibinfo {author}
  {\bibfnamefont {S.}~\bibnamefont {Wang}}, \bibinfo {author} {\bibfnamefont
  {H.}~\bibnamefont {Xu}}, \bibinfo {author} {\bibfnamefont {A.~G.}\
  \bibnamefont {Borisov}}, \bibinfo {author} {\bibfnamefont {H.}~\bibnamefont
  {Gu}}, \bibinfo {author} {\bibfnamefont {P.}~\bibnamefont {Nordlander}},
  \bibinfo {author} {\bibfnamefont {J.}~\bibnamefont {Aizpurua}}, \ and\
  \bibinfo {author} {\bibfnamefont {J.}~\bibnamefont {Ye}},\ }\href@noop {}
  {\bibfield  {journal} {\bibinfo  {journal} {Nano Lett.}\ }\textbf {\bibinfo
  {volume} {15}},\ \bibinfo {pages} {6419} (\bibinfo {year}
  {2015})}\BibitemShut {NoStop}%
\bibitem [{\citenamefont {Lassiter}\ \emph {et~al.}(2014)\citenamefont
  {Lassiter}, \citenamefont {Chen}, \citenamefont {Liu}, \citenamefont
  {Cirac{\`\i}}, \citenamefont {Hoang}, \citenamefont {Larouche}, \citenamefont
  {Oh}, \citenamefont {Mikkelsen},\ and\ \citenamefont
  {Smith}}]{Lassiter:2014gs}%
  \BibitemOpen
  \bibfield  {author} {\bibinfo {author} {\bibfnamefont {J.~B.}\ \bibnamefont
  {Lassiter}}, \bibinfo {author} {\bibfnamefont {X.}~\bibnamefont {Chen}},
  \bibinfo {author} {\bibfnamefont {X.}~\bibnamefont {Liu}}, \bibinfo {author}
  {\bibfnamefont {C.}~\bibnamefont {Cirac{\`\i}}}, \bibinfo {author}
  {\bibfnamefont {T.~B.}\ \bibnamefont {Hoang}}, \bibinfo {author}
  {\bibfnamefont {S.}~\bibnamefont {Larouche}}, \bibinfo {author}
  {\bibfnamefont {S.-H.}\ \bibnamefont {Oh}}, \bibinfo {author} {\bibfnamefont
  {M.~H.}\ \bibnamefont {Mikkelsen}}, \ and\ \bibinfo {author} {\bibfnamefont
  {D.~R.}\ \bibnamefont {Smith}},\ }\href@noop {} {\bibfield  {journal}
  {\bibinfo  {journal} {ACS Photonics}\ }\textbf {\bibinfo {volume} {1}},\
  \bibinfo {pages} {1212} (\bibinfo {year} {2014})}\BibitemShut {NoStop}%
\bibitem [{\citenamefont {Kheifets}\ \emph {et~al.}(2014)\citenamefont
  {Kheifets}, \citenamefont {Simha}, \citenamefont {Melin}, \citenamefont
  {Li},\ and\ \citenamefont {Raizen}}]{Kheifets:2014hq}%
  \BibitemOpen
  \bibfield  {author} {\bibinfo {author} {\bibfnamefont {S.}~\bibnamefont
  {Kheifets}}, \bibinfo {author} {\bibfnamefont {A.}~\bibnamefont {Simha}},
  \bibinfo {author} {\bibfnamefont {K.}~\bibnamefont {Melin}}, \bibinfo
  {author} {\bibfnamefont {T.}~\bibnamefont {Li}}, \ and\ \bibinfo {author}
  {\bibfnamefont {M.~G.}\ \bibnamefont {Raizen}},\ }\href@noop {} {\bibfield
  {journal} {\bibinfo  {journal} {Science}\ }\textbf {\bibinfo {volume}
  {343}},\ \bibinfo {pages} {1493} (\bibinfo {year} {2014})}\BibitemShut
  {NoStop}%
\bibitem [{\citenamefont {Chikkaraddy}\ \emph {et~al.}(6)\citenamefont
  {Chikkaraddy}, \citenamefont {de~Nijs}, \citenamefont {Benz}, \citenamefont
  {Barrow}, \citenamefont {Scherman}, \citenamefont {Rosta}, \citenamefont
  {Demetriadou}, \citenamefont {Fox}, \citenamefont {Hess},\ and\ \citenamefont
  {Baumberg}}]{Chikkaraddy:6ja}%
  \BibitemOpen
  \bibfield  {author} {\bibinfo {author} {\bibfnamefont {R.}~\bibnamefont
  {Chikkaraddy}}, \bibinfo {author} {\bibfnamefont {B.}~\bibnamefont
  {de~Nijs}}, \bibinfo {author} {\bibfnamefont {F.}~\bibnamefont {Benz}},
  \bibinfo {author} {\bibfnamefont {S.~J.}\ \bibnamefont {Barrow}}, \bibinfo
  {author} {\bibfnamefont {O.~A.}\ \bibnamefont {Scherman}}, \bibinfo {author}
  {\bibfnamefont {E.}~\bibnamefont {Rosta}}, \bibinfo {author} {\bibfnamefont
  {A.}~\bibnamefont {Demetriadou}}, \bibinfo {author} {\bibfnamefont
  {P.}~\bibnamefont {Fox}}, \bibinfo {author} {\bibfnamefont {O.}~\bibnamefont
  {Hess}}, \ and\ \bibinfo {author} {\bibfnamefont {J.~J.}\ \bibnamefont
  {Baumberg}},\ }\href@noop {} {\bibfield  {journal} {\bibinfo  {journal}
  {Nature}\ ,\ \bibinfo {pages} {1}} (\bibinfo {year} {6})}\BibitemShut
  {NoStop}%
\bibitem [{\citenamefont {Xiang}\ \emph {et~al.}(2014)\citenamefont {Xiang},
  \citenamefont {Zhang}, \citenamefont {Neuhauser},\ and\ \citenamefont
  {Lu}}]{Xiang:2014cz}%
  \BibitemOpen
  \bibfield  {author} {\bibinfo {author} {\bibfnamefont {H.}~\bibnamefont
  {Xiang}}, \bibinfo {author} {\bibfnamefont {X.}~\bibnamefont {Zhang}},
  \bibinfo {author} {\bibfnamefont {D.}~\bibnamefont {Neuhauser}}, \ and\
  \bibinfo {author} {\bibfnamefont {G.}~\bibnamefont {Lu}},\ }\href@noop {}
  {\bibfield  {journal} {\bibinfo  {journal} {J. Phys. Chem. Lett.}\ }\textbf
  {\bibinfo {volume} {5}},\ \bibinfo {pages} {1163} (\bibinfo {year}
  {2014})}\BibitemShut {NoStop}%
\bibitem [{\citenamefont {Xiang}\ \emph {et~al.}(2016)\citenamefont {Xiang},
  \citenamefont {Zhang}, \citenamefont {Zhang},\ and\ \citenamefont
  {Lu}}]{Xiang:2016cw}%
  \BibitemOpen
  \bibfield  {author} {\bibinfo {author} {\bibfnamefont {H.}~\bibnamefont
  {Xiang}}, \bibinfo {author} {\bibfnamefont {M.}~\bibnamefont {Zhang}},
  \bibinfo {author} {\bibfnamefont {X.}~\bibnamefont {Zhang}}, \ and\ \bibinfo
  {author} {\bibfnamefont {G.}~\bibnamefont {Lu}},\ }\href@noop {} {\bibfield
  {journal} {\bibinfo  {journal} {J. Phys. Chem. C}\ ,\ \bibinfo {pages}
  {acs.jpcc.6b05841}} (\bibinfo {year} {2016})}\BibitemShut {NoStop}%
\bibitem [{\citenamefont {Cirac{\`\i}}\ \emph
  {et~al.}(2013{\natexlab{b}})\citenamefont {Cirac{\`\i}}, \citenamefont
  {Pendry},\ and\ \citenamefont {Smith}}]{Ciraci:2013dz}%
  \BibitemOpen
  \bibfield  {author} {\bibinfo {author} {\bibfnamefont {C.}~\bibnamefont
  {Cirac{\`\i}}}, \bibinfo {author} {\bibfnamefont {J.~B.}\ \bibnamefont
  {Pendry}}, \ and\ \bibinfo {author} {\bibfnamefont {D.~R.}\ \bibnamefont
  {Smith}},\ }\href@noop {} {\bibfield  {journal} {\bibinfo  {journal}
  {ChemPhysChem}\ }\textbf {\bibinfo {volume} {14}},\ \bibinfo {pages} {1109}
  (\bibinfo {year} {2013}{\natexlab{b}})}\BibitemShut {NoStop}%
\bibitem [{\citenamefont {Raza}\ \emph {et~al.}(2015)\citenamefont {Raza},
  \citenamefont {Bozhevolnyi}, \citenamefont {Wubs},\ and\ \citenamefont
  {Mortensen}}]{Raza:2015ef}%
  \BibitemOpen
  \bibfield  {author} {\bibinfo {author} {\bibfnamefont {S.}~\bibnamefont
  {Raza}}, \bibinfo {author} {\bibfnamefont {S.~I.}\ \bibnamefont
  {Bozhevolnyi}}, \bibinfo {author} {\bibfnamefont {M.}~\bibnamefont {Wubs}}, \
  and\ \bibinfo {author} {\bibfnamefont {N.~A.}\ \bibnamefont {Mortensen}},\
  }\href@noop {} {\bibfield  {journal} {\bibinfo  {journal} {J. Phys.: Condens.
  Mat.}\ }\textbf {\bibinfo {volume} {27}},\ \bibinfo {pages} {183204}
  (\bibinfo {year} {2015})}\BibitemShut {NoStop}%
\bibitem [{\citenamefont {Toscano}\ \emph {et~al.}(2015)\citenamefont
  {Toscano}, \citenamefont {Straubel}, \citenamefont {Kwiatkowski},
  \citenamefont {Rockstuhl}, \citenamefont {Evers}, \citenamefont {Xu},
  \citenamefont {Mortensen},\ and\ \citenamefont {Wubs}}]{Toscano:2015iw}%
  \BibitemOpen
  \bibfield  {author} {\bibinfo {author} {\bibfnamefont {G.}~\bibnamefont
  {Toscano}}, \bibinfo {author} {\bibfnamefont {J.}~\bibnamefont {Straubel}},
  \bibinfo {author} {\bibfnamefont {A.}~\bibnamefont {Kwiatkowski}}, \bibinfo
  {author} {\bibfnamefont {C.}~\bibnamefont {Rockstuhl}}, \bibinfo {author}
  {\bibfnamefont {F.}~\bibnamefont {Evers}}, \bibinfo {author} {\bibfnamefont
  {H.}~\bibnamefont {Xu}}, \bibinfo {author} {\bibfnamefont {N.~A.}\
  \bibnamefont {Mortensen}}, \ and\ \bibinfo {author} {\bibfnamefont
  {M.}~\bibnamefont {Wubs}},\ }\href@noop {} {\bibfield  {journal} {\bibinfo
  {journal} {Nat. Comm.}\ }\textbf {\bibinfo {volume} {6}},\ \bibinfo {pages}
  {7132} (\bibinfo {year} {2015})}\BibitemShut {NoStop}%
\bibitem [{\citenamefont {Yan}(2015)}]{Yan:2015ff}%
  \BibitemOpen
  \bibfield  {author} {\bibinfo {author} {\bibfnamefont {W.}~\bibnamefont
  {Yan}},\ }\href@noop {} {\bibfield  {journal} {\bibinfo  {journal} {Phys.
  Rev. B}\ }\textbf {\bibinfo {volume} {91}},\ \bibinfo {pages} {115416}
  (\bibinfo {year} {2015})}\BibitemShut {NoStop}%
\bibitem [{\citenamefont {Cirac{\`\i}}\ and\ \citenamefont
  {Della~Sala}(2016)}]{Ciraci:2016il}%
  \BibitemOpen
  \bibfield  {author} {\bibinfo {author} {\bibfnamefont {C.}~\bibnamefont
  {Cirac{\`\i}}}\ and\ \bibinfo {author} {\bibfnamefont {F.}~\bibnamefont
  {Della~Sala}},\ }\href@noop {} {\bibfield  {journal} {\bibinfo  {journal}
  {Phys. Rev. B}\ }\textbf {\bibinfo {volume} {93}},\ \bibinfo {pages} {205405}
  (\bibinfo {year} {2016})}\BibitemShut {NoStop}%
\bibitem [{\citenamefont {Li}\ \emph {et~al.}(2015)\citenamefont {Li},
  \citenamefont {Fang}, \citenamefont {Weng}, \citenamefont {Zhang},
  \citenamefont {Dou}, \citenamefont {Yang},\ and\ \citenamefont
  {Yuan}}]{Li:2015io}%
  \BibitemOpen
  \bibfield  {author} {\bibinfo {author} {\bibfnamefont {X.}~\bibnamefont
  {Li}}, \bibinfo {author} {\bibfnamefont {H.}~\bibnamefont {Fang}}, \bibinfo
  {author} {\bibfnamefont {X.}~\bibnamefont {Weng}}, \bibinfo {author}
  {\bibfnamefont {L.}~\bibnamefont {Zhang}}, \bibinfo {author} {\bibfnamefont
  {X.}~\bibnamefont {Dou}}, \bibinfo {author} {\bibfnamefont {A.}~\bibnamefont
  {Yang}}, \ and\ \bibinfo {author} {\bibfnamefont {X.}~\bibnamefont {Yuan}},\
  }\href@noop {} {\bibfield  {journal} {\bibinfo  {journal} {Opt. Express}\
  }\textbf {\bibinfo {volume} {23}},\ \bibinfo {pages} {29738} (\bibinfo {year}
  {2015})}\BibitemShut {NoStop}%
\bibitem [{\citenamefont {Vignale}\ and\ \citenamefont
  {Kohn}(1996)}]{Vignale:1996hk}%
  \BibitemOpen
  \bibfield  {author} {\bibinfo {author} {\bibfnamefont {G.}~\bibnamefont
  {Vignale}}\ and\ \bibinfo {author} {\bibfnamefont {W.}~\bibnamefont {Kohn}},\
  }\href@noop {} {\bibfield  {journal} {\bibinfo  {journal} {Phys. Rev. Lett.}\
  }\textbf {\bibinfo {volume} {77}},\ \bibinfo {pages} {2037} (\bibinfo {year}
  {1996})}\BibitemShut {NoStop}%
\bibitem [{\citenamefont {Perdew}\ and\ \citenamefont
  {Zunger}(1981)}]{Perdew:1981dv}%
  \BibitemOpen
  \bibfield  {author} {\bibinfo {author} {\bibfnamefont {J.~P.}\ \bibnamefont
  {Perdew}}\ and\ \bibinfo {author} {\bibfnamefont {A.}~\bibnamefont
  {Zunger}},\ }\href@noop {} {\bibfield  {journal} {\bibinfo  {journal} {Phys.
  Rev. B}\ }\textbf {\bibinfo {volume} {23}},\ \bibinfo {pages} {5048}
  (\bibinfo {year} {1981})}\BibitemShut {NoStop}%
\bibitem [{\citenamefont {Vignale}\ \emph {et~al.}(1997)\citenamefont
  {Vignale}, \citenamefont {Ullrich},\ and\ \citenamefont
  {Conti}}]{Vignale:1997jda}%
  \BibitemOpen
  \bibfield  {author} {\bibinfo {author} {\bibfnamefont {G.}~\bibnamefont
  {Vignale}}, \bibinfo {author} {\bibfnamefont {C.~A.}\ \bibnamefont
  {Ullrich}}, \ and\ \bibinfo {author} {\bibfnamefont {S.}~\bibnamefont
  {Conti}},\ }\href@noop {} {\bibfield  {journal} {\bibinfo  {journal} {Phys.
  Rev. Lett.}\ }\textbf {\bibinfo {volume} {79}},\ \bibinfo {pages} {4878}
  (\bibinfo {year} {1997})}\BibitemShut {NoStop}%
\bibitem [{\citenamefont {Gross}\ and\ \citenamefont
  {Kohn}(1985)}]{Gross:1985gx}%
  \BibitemOpen
  \bibfield  {author} {\bibinfo {author} {\bibfnamefont {E.~K.~U.}\
  \bibnamefont {Gross}}\ and\ \bibinfo {author} {\bibfnamefont
  {W.}~\bibnamefont {Kohn}},\ }\href@noop {} {\bibfield  {journal} {\bibinfo
  {journal} {Phys. Rev. Lett.}\ }\textbf {\bibinfo {volume} {55}},\ \bibinfo
  {pages} {2850} (\bibinfo {year} {1985})}\BibitemShut {NoStop}%
\bibitem [{\citenamefont {Iwamoto}\ and\ \citenamefont
  {Gross}(1987)}]{Iwamoto:1987es}%
  \BibitemOpen
  \bibfield  {author} {\bibinfo {author} {\bibfnamefont {N.}~\bibnamefont
  {Iwamoto}}\ and\ \bibinfo {author} {\bibfnamefont {E.~K.~U.}\ \bibnamefont
  {Gross}},\ }\href@noop {} {\bibfield  {journal} {\bibinfo  {journal} {Phys.
  Rev. B}\ }\textbf {\bibinfo {volume} {35}},\ \bibinfo {pages} {3003}
  (\bibinfo {year} {1987})}\BibitemShut {NoStop}%
\bibitem [{\citenamefont {Nifosi}\ \emph {et~al.}(1998)\citenamefont {Nifosi},
  \citenamefont {Conti},\ and\ \citenamefont {Tosi}}]{Nifosi:1998fa}%
  \BibitemOpen
  \bibfield  {author} {\bibinfo {author} {\bibfnamefont {R.}~\bibnamefont
  {Nifosi}}, \bibinfo {author} {\bibfnamefont {S.}~\bibnamefont {Conti}}, \
  and\ \bibinfo {author} {\bibfnamefont {M.~P.}\ \bibnamefont {Tosi}},\
  }\href@noop {} {\bibfield  {journal} {\bibinfo  {journal} {Phys. Rev. B}\
  }\textbf {\bibinfo {volume} {58}},\ \bibinfo {pages} {12758} (\bibinfo {year}
  {1998})}\BibitemShut {NoStop}%
\bibitem [{\citenamefont {Qian}\ and\ \citenamefont
  {Vignale}(2002)}]{Qian:2002dl}%
  \BibitemOpen
  \bibfield  {author} {\bibinfo {author} {\bibfnamefont {Z.}~\bibnamefont
  {Qian}}\ and\ \bibinfo {author} {\bibfnamefont {G.}~\bibnamefont {Vignale}},\
  }\href@noop {} {\bibfield  {journal} {\bibinfo  {journal} {Phys. Rev. B}\
  }\textbf {\bibinfo {volume} {65}},\ \bibinfo {pages} {235121} (\bibinfo
  {year} {2002})}\BibitemShut {NoStop}%
\bibitem [{\citenamefont {Conti}\ and\ \citenamefont
  {Vignale}(1999)}]{Conti:1999fv}%
  \BibitemOpen
  \bibfield  {author} {\bibinfo {author} {\bibfnamefont {S.}~\bibnamefont
  {Conti}}\ and\ \bibinfo {author} {\bibfnamefont {G.}~\bibnamefont
  {Vignale}},\ }\href@noop {} {\bibfield  {journal} {\bibinfo  {journal} {Phys.
  Rev. B}\ }\textbf {\bibinfo {volume} {60}},\ \bibinfo {pages} {7966}
  (\bibinfo {year} {1999})}\BibitemShut {NoStop}%
\bibitem [{Note1()}]{Note1}%
  \BibitemOpen
  \bibinfo {note} {\protect \textit {Supplemental Material}}\BibitemShut
  {NoStop}%
\bibitem [{\citenamefont {Giugliani}\ and\ \citenamefont
  {Vignale}(2005)}]{Giuliani:2005ve}%
  \BibitemOpen
  \bibfield  {author} {\bibinfo {author} {\bibfnamefont {G.}~\bibnamefont
  {Giugliani}}\ and\ \bibinfo {author} {\bibfnamefont {G.}~\bibnamefont
  {Vignale}},\ }\href@noop {} {\emph {\bibinfo {title} {{Quantum Theory of the
  Electron Liquid}}}}\ (\bibinfo  {publisher} {Cambridge University Press},\
  \bibinfo {year} {2005})\BibitemShut {NoStop}%
\bibitem [{Note2()}]{Note2}%
  \BibitemOpen
  \bibinfo {note} {\protect \textsc {Comsol} Multiphysics,
  http://www.comsol.com}\BibitemShut {NoStop}%
\bibitem [{\citenamefont {Cirac{\`\i}}\ \emph
  {et~al.}(2013{\natexlab{c}})\citenamefont {Cirac{\`\i}}, \citenamefont
  {Urzhumov},\ and\ \citenamefont {Smith}}]{Ciraci:2013wi}%
  \BibitemOpen
  \bibfield  {author} {\bibinfo {author} {\bibfnamefont {C.}~\bibnamefont
  {Cirac{\`\i}}}, \bibinfo {author} {\bibfnamefont {Y.~A.}\ \bibnamefont
  {Urzhumov}}, \ and\ \bibinfo {author} {\bibfnamefont {D.~R.}\ \bibnamefont
  {Smith}},\ }\href@noop {} {\bibfield  {journal} {\bibinfo  {journal} {Opt.
  Express}\ }\textbf {\bibinfo {volume} {21}},\ \bibinfo {pages} {9397}
  (\bibinfo {year} {2013}{\natexlab{c}})}\BibitemShut {NoStop}%
\bibitem [{\citenamefont {Kreibig}\ and\ \citenamefont
  {Vollmer}(2013)}]{Kreibig:2013ci}%
  \BibitemOpen
  \bibfield  {author} {\bibinfo {author} {\bibfnamefont {U.}~\bibnamefont
  {Kreibig}}\ and\ \bibinfo {author} {\bibfnamefont {M.}~\bibnamefont
  {Vollmer}},\ }\href@noop {} {\emph {\bibinfo {title} {{Optical Properties of
  Metal Clusters}}}},\ \bibinfo {series} {Springer Series in Materials
  Science}, Vol.~\bibinfo {volume} {25}\ (\bibinfo  {publisher} {Springer
  Science {\&} Business Media},\ \bibinfo {address} {Berlin, Heidelberg},\
  \bibinfo {year} {2013})\BibitemShut {NoStop}%
\bibitem [{\citenamefont {Barbry}\ \emph {et~al.}(2015)\citenamefont {Barbry},
  \citenamefont {Koval}, \citenamefont {Marchesin}, \citenamefont {Esteban},
  \citenamefont {Borisov}, \citenamefont {Aizpurua},\ and\ \citenamefont
  {S{\'a}nchez-Portal}}]{Barbry:2015iw}%
  \BibitemOpen
  \bibfield  {author} {\bibinfo {author} {\bibfnamefont {M.}~\bibnamefont
  {Barbry}}, \bibinfo {author} {\bibfnamefont {P.}~\bibnamefont {Koval}},
  \bibinfo {author} {\bibfnamefont {F.}~\bibnamefont {Marchesin}}, \bibinfo
  {author} {\bibfnamefont {R.}~\bibnamefont {Esteban}}, \bibinfo {author}
  {\bibfnamefont {A.~G.}\ \bibnamefont {Borisov}}, \bibinfo {author}
  {\bibfnamefont {J.}~\bibnamefont {Aizpurua}}, \ and\ \bibinfo {author}
  {\bibfnamefont {D.}~\bibnamefont {S{\'a}nchez-Portal}},\ }\href@noop {}
  {\bibfield  {journal} {\bibinfo  {journal} {Nano Lett.}\ }\textbf {\bibinfo
  {volume} {15}},\ \bibinfo {pages} {3410} (\bibinfo {year}
  {2015})}\BibitemShut {NoStop}%
\bibitem [{\citenamefont {Mortensen}\ \emph {et~al.}(2014)\citenamefont
  {Mortensen}, \citenamefont {Raza}, \citenamefont {Wubs}, \citenamefont
  {S{\o}ndergaard},\ and\ \citenamefont {Bozhevolnyi}}]{Mortensen:2014kc}%
  \BibitemOpen
  \bibfield  {author} {\bibinfo {author} {\bibfnamefont {N.~A.}\ \bibnamefont
  {Mortensen}}, \bibinfo {author} {\bibfnamefont {S.}~\bibnamefont {Raza}},
  \bibinfo {author} {\bibfnamefont {M.}~\bibnamefont {Wubs}}, \bibinfo {author}
  {\bibfnamefont {T.}~\bibnamefont {S{\o}ndergaard}}, \ and\ \bibinfo {author}
  {\bibfnamefont {S.~I.}\ \bibnamefont {Bozhevolnyi}},\ }\href@noop {}
  {\bibfield  {journal} {\bibinfo  {journal} {Nat. Comm.}\ }\textbf {\bibinfo
  {volume} {5}},\ \bibinfo {pages} {3809} (\bibinfo {year} {2014})}\BibitemShut
  {NoStop}%
\end{thebibliography}
\end{document}